\documentclass[journal]{IEEEtranTIE}
\usepackage{graphicx}
\usepackage{cite}
\usepackage{picinpar}
\usepackage{amsmath}
\usepackage{amssymb}
\usepackage{url}
\usepackage{flushend}
\usepackage[utf8]{inputenc}
\usepackage{colortbl}
\usepackage{soul}
\usepackage{multirow}
\usepackage{pifont}
\usepackage{color}
\usepackage{alltt}
\usepackage[hidelinks]{hyperref}
\usepackage{enumerate}
\usepackage{siunitx}
\usepackage{breakurl}
\usepackage{epstopdf}
\usepackage{pbox}
\usepackage{algorithm}
\usepackage{algorithmic}

\begin{document}
\title{	Mixtac: A Novel Bio-Inspired Hybrid Tactile Sensor with Synergistic Event-Frame Perception}

\author{
	\vskip 1em
	
	Yihang~Li,
	Yijin~Chen,
	Junkai~Xu,
	Na~Ningguta,
	Peter~B.~Shull, \emph{Member, IEEE},
	Shuo~Jiang, \emph{Member, IEEE},
	\\ and Bin~He, \emph{Senior Member, IEEE}

	\thanks{
		This work was supported in part by the National Key R\&D Program of China under Grant 2024YFB4707600, in part by the National Natural Science Foundation of China under Grant 62573321 and Grant W2441018, and in part by the Shanghai Municipal Commission of Science and Technology under Grant 24511104400.

		Yihang Li, Yijin Chen, Shuo Jiang, and Bin He are with the Shanghai Research Institute for Intelligent Autonomous Systems, the State Key Laboratory of Autonomous Intelligent Unmanned Systems, and the College of Electronics and Information Engineering, Tongji University, Shanghai 201804, China (e-mail: jiangshuo@tongji.edu.cn).

		Peter B. Shull is with the State Key Laboratory of Mechanical System and Vibration, School of Mechanical Engineering, Shanghai Jiao Tong University, Shanghai 200240, China (e-mail: pshull@sjtu.edu.cn).

		Shuo Jiang is the corresponding author (e-mail: jiangshuo@tongji.edu.cn).
	}
}

\maketitle
	
\begin{abstract}
Vision based and event based tactile sensors are important in robotic manipulation research. However, they suffer from a fundamental tradeoff: vision based sensors have low sampling rates, while event based sensors are prone to drift during long term static force estimation. To solve this challenge and achieve human level tactile perception, the novel hybrid event frame tactile sensor (Mixtac) is proposed in this paper by emulating the synergistic function of biological mechanoreceptors, which achieves normal force estimation. The prototype leverages events for high frequency force tracking and frames for long term accuracy. The Frame Guided Event Recurrent Network (FGER-Net) was proposed to fuse the two data streams. Frames were used by the net to correct event drift during training and guide high frequency predictions during inference. Experiments demonstrated an MAE of 0.04 N. This paper could bridge the sampling rate gap from 0 to 500 Hz in current vision based tactile sensors and pave the way for human level robotic manipulation.
\end{abstract}

\begin{IEEEkeywords}
Deep learning, event camera, force estimation, sensor fusion, vision-based tactile sensor.
\end{IEEEkeywords}

\markboth{IEEE TRANSACTIONS ON INDUSTRIAL ELECTRONICS}%
{}

\definecolor{limegreen}{rgb}{0.2, 0.8, 0.2}
\definecolor{forestgreen}{rgb}{0.13, 0.55, 0.13}
\definecolor{greenhtml}{rgb}{0.0, 0.5, 0.0}

\section{Introduction}

\IEEEPARstart{T}{he} ability for a robot to perceive and respond to physical contact underpins dexterous object manipulation\cite{liu2023enhancing,hogan2020tactile}. In particular, achieving human level dexterous performance requires the perception of both rapidly changing dynamic forces and stable static forces\cite{yousef2011tactile}. Among them, vision-based tactile sensors are widely adopted due to their high integration and compatibility with computer vision pipelines~\cite{zhangHardwareTechnologyVisionBased2022,hogan2020tactile}.
Accurate perception of dynamic force changes enables rapid adjustment of fingertip force and posture, thereby supporting stable manipulation and preventing slip or damage~\cite{luo2017robotic}. Conversely, delayed perception can hinder timely control responses and ultimately lead to grasp failure. Therefore, accurate measurement of dynamic force signals requires a high sampling rate\cite{dwivedi2018design}. In particular, emulating human tactile vibration perception demands a sampling rate of at least 400 Hz to reliably detect high frequency~\cite{tiwana2012review}. Beyond perception, high sampling rates are essential for stable closed-loop force control, as low sampling rates introduce delays that compromise passivity and reduce responsiveness~\cite{romano2011human}.\par
However, achieving such high sampling rates with vision-based tactile sensors remains difficult without sacrificing compactness. Most tactile sensors based on vision rely on frame cameras, yet frame rate and computational constraints preclude the capture of signals at high frequencies \cite{naeini2019novel}. Moreover, ultra high speed frame cameras require large aperture optics to collect sufficient light at short exposure times, making compact integration difficult~\cite{miyashita2022portable}. \par
Meanwhile, robotic manipulation requires tactile sensors to provide stable force estimation over extended periods. Many tasks last several seconds from initial contact to completion, making long term stability essential.\par
These competing requirements create a fundamental tradeoff in vision based tactile sensing. Frame based sensors support stable static force estimation through rich and absolute spatial information. However, limited frame rates restrict dynamic sensing. Event based sensors capture rapid dynamics with microsecond temporal resolution, while their change based measurements are relative and prone to drift in static or prolonged scenarios.\par
Recently, event based tactile methods have shown clear potential for high frequency sensing and slip detection. However, stable long term absolute force estimation remains difficult due to event signals mainly encoding changes in deformation. Funk et al.~\cite{funk2024evetac} achieved tangential force estimation, but could not estimate normal force. Mukashev et al.~\cite{mukashevEBTSEventBasedTactile2025} achieved stable force estimation at 500 Hz by using active flickering illumination to generate continuous events. However, this strategy depends on high frequency light modulation. As a result, it increases data redundancy and hardware complexity, much like high speed frame based approaches. Yin et al.~\cite{yin2025gelevent} demonstrated high frequency force estimation with event cameras at 180 Hz. However, this method estimates force changes rather than absolute force, which limits long term stability. In addition, event cameras respond primarily to brightness changes. They therefore produce sparse signals when visual changes are small, such as during static grasping~\cite{he2024microsaccade}. Long term integration can also accumulate errors and degrade estimation accuracy over time~\cite{baghaei2020neuromorphic}. These limitations show that existing event based methods still require further improvement for tasks such as static grasping.\par
Fusion of frame images and event signals offers a promising way to address these challenges. Frame images and event signals provide complementary information for force estimation. Frame images preserve detailed spatial deformation, whereas event signals capture the rate of deformation. However, fusing the two modalities remains difficult. Parallax causes the two cameras to observe silicone deformation from different viewpoints and with different spatial offsets, which complicates spatial alignment and hybridization~\cite{wasti2024spatiotemporal}. In addition, the ultra high temporal resolution of event signals creates a major synchronization challenge. The mismatch between asynchronous events and synchronous frames further complicates effective fusion for force estimation~\cite{chen2024retain}. \par

\begin{figure*}
    \centering
    \includegraphics[width=7.16in]{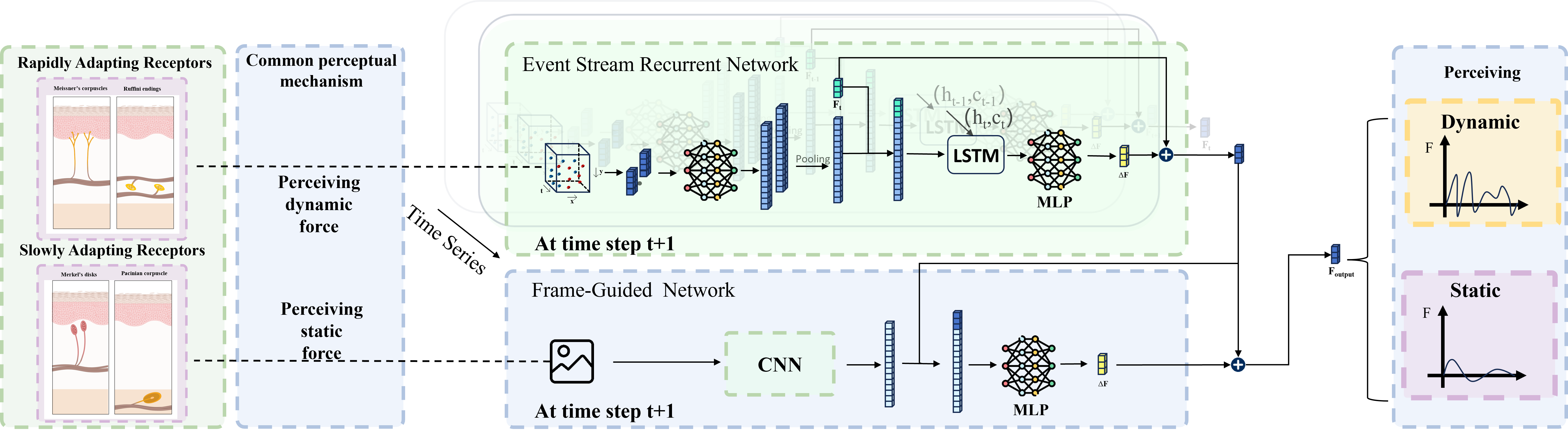}
    \caption{The architecture of the proposed FGER-Net is presented, alongside its conceptual underpinnings. The left portion illustrates the biological inspiration for the network: the synergistic function of fast and slow adapting mechanoreceptors in perceiving dynamic and static forces, respectively. These biological principles are emulated by the FGER-Net. In the network itself (center), event streams and image frames are synergistically fused for robust force estimation. In this architecture, frames are used to periodically correct for drift during training and to continuously stabilize the high frequency force predictions derived from the event stream during inference. The resulting force estimation is thereby made accurate and robust against drift, successfully capturing both dynamic and static tactile information, mirroring the perceptual capabilities shown on the right side of the diagram. }
    \label{fig_suanfa}
\end{figure*}

A potential solution to this challenge lies in the organization of biological tactile perception. Comprehensive human level tactile perception relies on the complementary functions of different neural receptors. Slow adapting receptors provide stable and persistent signals for static forces. Fast adapting receptors respond selectively to transient dynamic events. Pacinian corpuscles are especially sensitive to vibration and have a bandwidth of about 250 Hz~\cite{tiwana2012review}. Based on this biological inspiration, Mixtac is proposed to address the tradeoff between dynamic sensing and long term stability. FGER-Net is further developed to fuse event and frame signals for force estimation (Fig. \ref{fig_suanfa}).\par
In summary, the main contributions are as follows:\par

\begin{figure}[!t]
\centering

\includegraphics[width=3.5in]{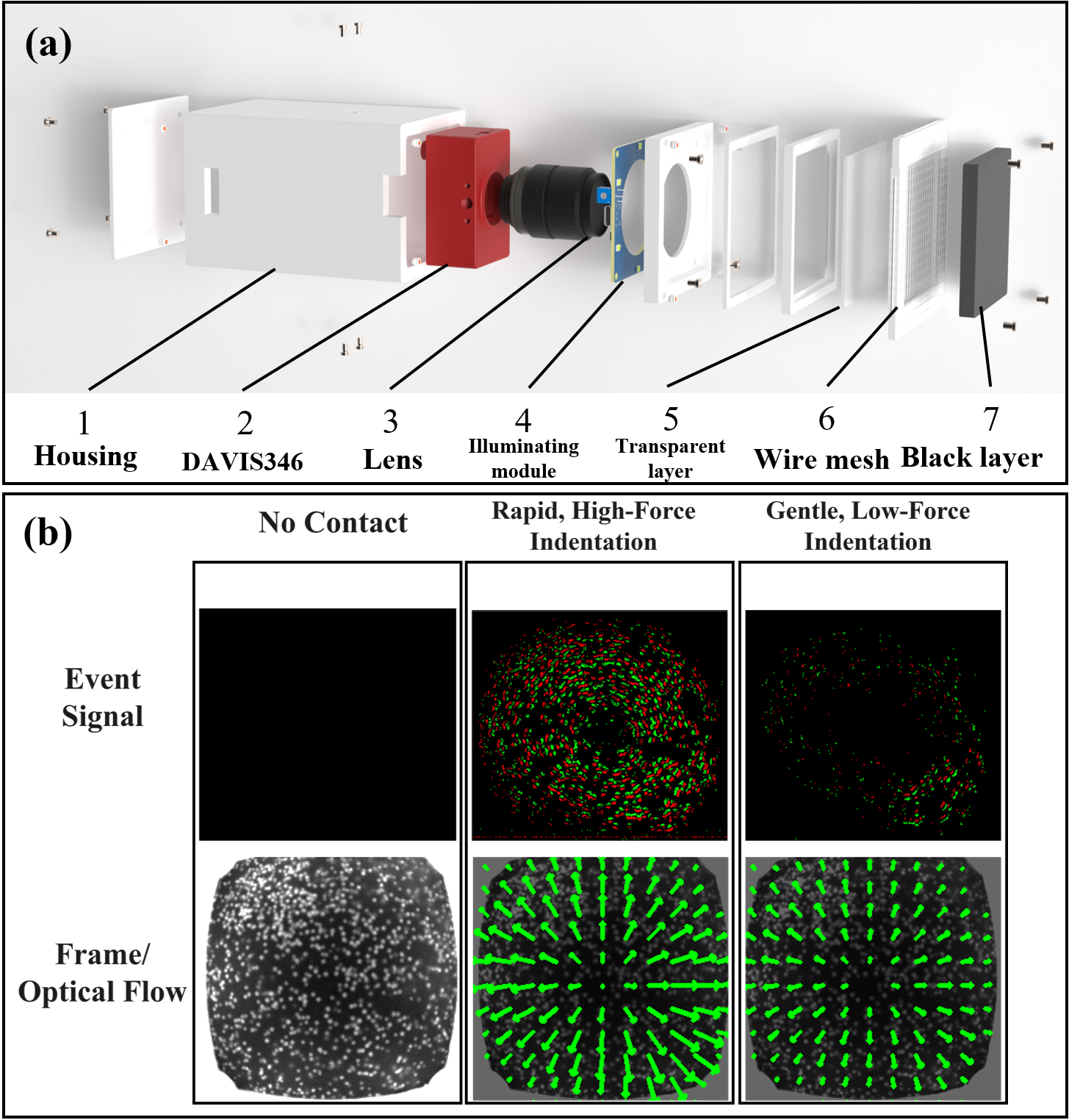}

\caption{Hardware design and sensor responses of Mixtac. (a) The overall structural layout of the hardware. (b) A comparison of the sensor’s response to varying indentation forces. From left to right, the panels demonstrate three states: no contact; rapid, high force indentation (resulting in dense event signals and a high magnitude optical flow); and gentle, low force indentation (resulting in sparse event signals and a low magnitude optical flow).
}
\label{fig_1}
\end{figure}

\begin{enumerate}

    \item A hybrid tactile sensor named Mixtac was designed and implemented, which can synergistically fuse event and frame data to achieve both high dynamic response and long term estimation stability (Fig. \ref{fig_1}(a)).
    \item The FGER-Net, a novel deep learning architecture for hybrid tactile data was developed. The network is designed to dynamically fuse the two data streams, learning to prioritize the event stream for tracking high-frequency dynamics while primarily relying on the frame stream to ensure stability and accuracy during static force estimation, which achieves normal force estimation.
    \item Extensive experimental validations were performed, demonstrating the system's synergistic fusion capability. The sensor achieves high fidelity dynamic force estimation and long term force stability (over 25 s). It also robustly handles hybrid force profiles that transition from transient impact to static hold through dynamic weighting of event and frame contributions.
 
\end{enumerate}

\begin{table}[!t]

\caption{Mixtac Hardware Components\label{tab:hardware}}
\centering

\begin{tabular}{cp{5.5cm}}
\hline
\textbf{Component} & \textbf{Specifications} \\
\hline
Camera & DAVIS346 from Inivation, 346 x 260 px resolution. \\[0.5em]
Housing & Custom Design, 3D printed. \\[0.5em]
Camera Screws & 1/2.8 in, thread length 2.8 mm. \\[0.5em]
Light module & Custom PCB, LED model XL-2835UWC-02. \\[0.5em]
Contact module &
    Shore A-5 for transparent layer. 
    Shore A-5 for black layer. 
    Wire mesh, aperture 0.9 mm, wire diameter 0.18 mm. \\[0.5em]
Marker & Glass microspheres, diameter 0.8 mm. \\[0.5em]
\hline
\end{tabular}
\end{table}

\section{RELATED WORK}

\subsection{Frame Based Optical Tactile Sensor}
Frame based optical tactile sensing has been widely studied. It offers high integration, strong interference immunity. Representative systems include TacTip~\cite{ward2018tactip}. Recent work has improved tactile reconstruction and force perception. GelStereo enabled submillimeter tactile point cloud perception for localization and small part insertion~\cite{9464700}. Other studies expanded the sensing functions of optical tactile devices. Nozaki and Krebs proposed an optical sensor that measures distance, surface tilt, and contact force~\cite{9446588}. GelStereo 2.0 improved tactile reconstruction under refractive imaging conditions~\cite{10255368}. Compact optical force sensing has also drawn attention. Mo et al. presented a compact soft triaxial optical force sensor for robotic grasping and human robot interaction~\cite{mo2025highly}. Even so, effective force sampling rates usually remain within 10 to 200 Hz. This is still below the bandwidth required for human level high frequency tactile perception, which is about 400 Hz~\cite{tiwana2012review}.\par

\subsection{Event Based Optical Tactile Sensor}

Event based optical tactile sensing is regarded as a promising route for high sampling perception due to high temporal resolution and low latency response~\cite{naeini2019novel}. Update rates above 1 kHz have been reported, but force observability is often incomplete (e.g., shear dominant estimation without reliable normal force recovery)~\cite{funk2024evetac}, and active flickering illumination is frequently required, which weakens the low redundancy advantage of event driven sensing in nearly static conditions~\cite{mukashevEBTSEventBasedTactile2025}. Three-dimensional force estimation has also been demonstrated at moderate rates (e.g., 180 Hz), yet the reported contact duration remains short (0.25-1 s)~\cite{yin2025gelevent}. Although no explicit cause was provided in the original study, this limitation is plausibly related to sparse responses in weak deformation intervals and error accumulation during long horizon integration of differential event signals.
\par

\section{METHODOLOGY}

\subsection{Hardware Components}
\subsubsection{Contacting Module}

The proposed design paradigm for a compact self supporting sensor is realized in the architecture of the contacting module. In this design, the rigid acrylic support plate is eliminated, and support is instead provided by a flexible steel mesh embedded directly within the silicone elastomer. Through this embedded support strategy, the thickness of the contacting module is significantly reduced, while the required supporting effect is maintained.\par

The module is designed to provide stable and information rich deformation cues for force reconstruction. It consists of a transparent sensing silicone layer with randomly dispersed glass microspheres and a black protective layer for ambient light shielding. The 3D microsphere distribution encodes depth dependent strain, while a thickening agent is incorporated to prevent marker sedimentation during curing. In contrast to conventional designs that rely on a rigid acrylic backplate, our sensor embeds a flexible steel mesh support layer within the elastomer to improve structural compliance. \par

To suppress reflection induced noise in event sensing, the inner surface of the transparent silicone is treated by high grit micro sanding. 
Unlike the coarse anti reflection sanding used in our early acrylic backed prototypes, which caused marker blur, this finer treatment effectively reduces specular reflection while preserving spatial detail. 
Together, these design choices provide a clearer and higher contrast marker image for robust event frame perception. A random marker layout enhances event responses during deformation. Compared with regular grids, dense random markers provide more motion cues and improve observability of fine-grained dynamics.

Multi step fabrication process is further used to ensure a uniform 3D distribution of glass microbeads in the silicone matrix. Marker settling caused by density mismatch is mitigated by increasing the viscosity of liquid silicone with a thickening agent, and the resulting degassing difficulty is addressed through mechanical stirring followed by dual stage vacuum pumping. As a result, a transparent and bubble free sensing elastomer is obtained.

\subsubsection{Illuminating Module}

Stable and uniform lighting is provided by a custom PCB that integrates XL-2835UWC-02 LEDs and follows the design principles established in \cite{lin20239dtact}. A CN5711 driver chip is integrated on the board to ensure stable, adjustable current. The entire component is powered by a standard 5V USB port. \par

\subsubsection{DAVIS346 Event-Frame Camera}

The DAVIS346 camera simultaneously outputs APS intensity frames and DVS event streams~\cite{taverni2018front}. Under different contact conditions (Fig.~\ref{fig_1}(b)), event density varies with indentation speed and captures high-frequency dynamics, whereas frame-based optical flow reflects deformation magnitude and provides absolute geometric information.\par

\subsection{Signal Processing}

\subsubsection{Calibrating the Camera}

To remove the geometric distortion from the wide angle lens and obtain accurate camera images, the camera's internal parameters were calibrated. The calibration process based on a standard checkerboard was adopted, which was implemented using the accompanying DV software.

\IEEEpubidadjcol

\subsubsection{Aligning Data}

Precise temporal alignment between event data and force labels is achieved using a software based calibration with mechanical impulses. In the absence of hardware synchronization, a sharp impulse is applied as a shared temporal marker, producing a force peak and an intermittent burst pattern. The event rate trough near transient silicone stillness at maximum deformation is treated as the event side anchor and paired with the force peak; these two timestamps correspond to the same physical contact state.

Two impulses at the beginning and end of each trial provide two anchor pairs, $(T_{E1}, T_{F1})$ and $(T_{E2}, T_{F2})$, where $T_{E*}$ and $T_{F*}$ denote event trough and force-peak timestamps, respectively. Assuming a linear clock relation, the mapping parameters ($a$, $b$) are solved from:
\begin{subequations}
\label{eq:system_solve}
\begin{align}
    T_{F1} &= a \cdot T_{E1} + b \label{eq:sys1} \\
    T_{F2} &= a \cdot T_{E2} + b \label{eq:sys2}
\end{align}
\end{subequations}
The solved $a$ (clock drift) and $b$ (time offset) are then used to map all event timestamps into the force sensor time domain:
\begin{equation}
\label{eq:time_alignment}
t_{\text{force}} = a \cdot t_{\text{event}} + b
\end{equation}
This transformation ensures temporal consistency for all subsequent samples within each trial.\par

Ground truth normal force is provided by a stationary 6 axis reference sensor. During data collection, Mixtac is rigidly mounted on the reference sensor, and contact is applied using a rigid hemispherical indenter. Owing to the coaxial alignment and rigid coupling between the two sensors, the measured z axis force is treated as the normal contact force, with lateral shear components effectively decoupled.\par

\subsubsection{Frame Guided Event Recurrent Network}

Mixtac's core algorithmic challenge lies in fusing high rate asynchronous events with low rate synchronous frames. Naive frame rate concatenation inherently discards critical temporal dynamics between frames. The fundamental tradeoff is clear: images provide spatially rich absolute deformation data but at insufficient sampling rates, while events provide microsecond level tracking of relative changes but inevitably accumulate drift over time.\par
Inspired by the human tactile system's synergy of fast and slow adapting mechanoreceptors, we propose the Frame Guided Event Recurrent Force Network (FGER-Net). This dual rate architecture mirrors biological perception: a high frequency recurrent pathway integrates transient force increments from the event stream, while a low frequency pathway extracts absolute spatial references from single image frames for periodic training corrections.\par

During inference, we maximize estimation stability via a continuous correction strategy. By persistently fusing the most recent frame feature with incoming events, this absolute spatial anchor effectively eliminates integration drift.\par

The architecture utilizes dual parallel pathways optimized for each modality's distinct characteristics. To circumvent the extreme sparsity typical of standard convolutions on event data, individual events are normalized and processed by a pointwise MLP, followed by max-pooling to generate an aggregated feature vector for each packet. Conversely, dense spatial information from image frames is extracted into a compact feature vector using a residual CNN backbone coupled with average pooling. \par

\paragraph{Data Stream Segmentation}
The algorithm's processing cycle is driven by the arrival of new frames. A cycle is defined as the time interval between two consecutive frames, $I_{k-1}$ and $I_k$. All events occurring within this interval are segmented into a sequence of event packets $P_m$ using a fixed time window $\Delta T$. Therefore, the number of event packets, M, depending on the camera's frame rate. 
\begin{multline}
    P_m = \{e_k \in E_{\text{stream}} \mid (m-1)\Delta T \le t_k < m\Delta T \}, \\
    \text{for } m=1, \dots, M
\end{multline}
where $P_m$ is the $m$-th event packet. 

\begin{algorithm}[t]
\caption{FGER-Net}\label{alg:fger_net}
\begin{algorithmic}
\STATE \textbf{Input:} $I_{\text{guide}}$, $\{P_m\}_{m=1}^M$, $f_{\text{prev\_cycle}}$, $(h_{\text{prev\_cycle}}, C_{\text{prev\_cycle}})$
\STATE \textbf{Output:} $\hat{f}_{\text{final}}$, $(h_M, C_M)$
\STATE \textit{Guiding Frame Feature Extraction}
\STATE $\mathbf{f}_{\text{frame}} \gets \Phi_f(I_{\text{guide}})$
\STATE \textit{Recurrent Processing Loop}
\STATE $\hat{f}_0 \gets f_{\text{prev\_cycle}}$; $(h_0, C_0) \gets (h_{\text{prev\_cycle}}, C_{\text{prev\_cycle}})$
\STATE \textbf{for} $m \gets 1$ \textbf{to} $M$ \textbf{do}
\STATE \hspace{0.5cm} $\mathbf{f}_{e,m} \gets \Phi_e(P_m)$
\STATE \hspace{0.5cm} $\mathbf{x}_m \gets \text{concatenate}(\mathbf{f}_{e,m}, \hat{f}_{m-1})$
\STATE \hspace{0.5cm} $(h_m, C_m) \gets \text{LSTM}(\mathbf{x}_m, (h_{m-1}, C_{m-1}))$
\STATE \hspace{0.5cm} $\Delta\hat{f}_m \gets \text{MLP}_{\Delta f}(h_m)$
\STATE \hspace{0.5cm} $\hat{f}_m \gets \hat{f}_{m-1} + \Delta\hat{f}_m$
\STATE \textbf{end for}
\STATE \textit{Fusion and Correction}
\STATE $\mathbf{f}_{\text{fusion}} \gets \text{concatenate}(\hat{f}_M, \mathbf{f}_{\text{frame}})$
\STATE $\Delta f_{\text{corr}} \gets \text{MLP}_{\text{corr}}(\mathbf{f}_{\text{fusion}})$
\STATE $\hat{f}_{\text{final}} \gets \hat{f}_M + \Delta f_{\text{corr}}$
\STATE \textbf{return} $\hat{f}_{\text{final}}, (h_M, C_M)$
\end{algorithmic}
\end{algorithm}

\paragraph{Guiding Frame Feature Extraction}
First, a CNN is used, $\Phi_f$ (e.g. a part of ResNet), to extract a high-dimensional spatial feature vector $f_{\text{frame}}$ from the guide frame $I_{\text{guide}}$. This feature remains constant throughout the recurrent estimation process for one cycle. 
\begin{equation}
   \mathbf{f_{\text{frame}}} = \Phi_f(I_{\text{guide}}) \in \mathbb{R}^{D_f}
\end{equation}
where $D_f$ is the dimension of the frame feature. 

\paragraph{Recurrent Processing Loop}
We initialize the scalar force estimate $\hat{f}_0$ and the LSTM states $(h_0, C_0)$ as zeros. Subsequently, for each event packet $P_m$ (where $m=1, \dots, M$), the network executes the following recurrent updates:

\begin{enumerate}
    \item \textbf{Event Feature Extraction:} A dedicated CNN branch, $\Phi_e$, extracts a high-dimensional temporal feature vector $\mathbf{f}_{e, m}$ from the current event packet $P_m$:
    \begin{equation}
        \mathbf{f}_{e, m} = \Phi_e(P_m) \in \mathbb{R}^{D_e}
    \end{equation}

    \item \textbf{Input Vector Construction:} The extracted event feature $\mathbf{f}_{e, m}$ is concatenated with the scalar force estimate from the previous step, $\hat{f}_{m-1}$, forming the input vector $\mathbf{x}_m$ for the LSTM module:
    \begin{equation}
        \mathbf{x}_m = [\mathbf{f}_{e, m} ; \hat{f}_{m-1}] \in \mathbb{R}^{D_e + 1}
    \end{equation}

    \item \textbf{LSTM Update:} The input vector $\mathbf{x}_m$ and the previous hidden states $(h_{m-1}, C_{m-1})$ are fed into the LSTM cell to compute the updated states $(h_m, C_m)$:
    \begin{equation}
        (h_m, C_m) = \text{LSTM}(\mathbf{x}_m, (h_{m-1}, C_{m-1}))
    \end{equation}
    
    \item \textbf{Force Increment Prediction:} A Multi-Layer Perceptron, denoted as $\text{MLP}_{\Delta f}$, regresses the high-frequency normal force increment $\Delta \hat{f}_m$ for the current time window derived from the hidden state $h_m$:
    \begin{equation}
        \Delta \hat{f}_m = \text{MLP}_{\Delta f}(h_m) \in \mathbb{R}
    \end{equation}

    \item \textbf{Force Accumulation:} The current normal force estimate is updated sequentially. This intermediate prediction relies purely on the integration of the fast-adapting event stream:
    \begin{equation}
        \hat{f}_m = \hat{f}_{m-1} + \Delta \hat{f}_m
    \end{equation}
\end{enumerate}

After iterating through all $M$ packets within the cycle, an accumulated normal force estimate $\hat{f}_M$, driven purely by the transient event stream, is obtained. 

\paragraph{Fusion Feature Construction}
To counteract integration drift, we concatenate the final accumulated event driven force, $\hat{f}_M$, with the absolute spatial guide frame feature, $\mathbf{f}_{\text{frame}}$, constructing a unified multimodal fusion vector, $\mathbf{f}_{\text{fusion}}$:
\begin{equation}
    \mathbf{f}_{\text{fusion}} = [\hat{f}_M ; \mathbf{f}_{\text{frame}}] \in \mathbb{R}^{1 + D_f}
\end{equation}

\paragraph{Correction Term Prediction}
The fusion feature $\mathbf{f}_{\text{fusion}}$ is processed by calibration network ($\text{MLP}_{\text{corr}}$) to predict a 1D force correction term, $\Delta f_{\text{corr}}$. This component explicitly learns to compensate for accumulated drift by anchoring the estimate to the reliable absolute spatial deformation captured by the image frame:
\begin{equation}
    \Delta f_{\text{corr}} = \text{MLP}_{\text{corr}}(\mathbf{f}_{\text{fusion}}) \in \mathbb{R}
\end{equation}

\paragraph{Final Force Output}
The corrective term is seamlessly added to the accumulated force to yield the final, drift compensated normal force estimate for the current temporal cycle, $\hat{f}_{\text{final}}$:
\begin{equation}
    \hat{f}_{\text{final}} = \hat{f}_M + \Delta f_{\text{corr}}
\end{equation}
This final prediction $\hat{f}_{\text{final}}$ subsequently initializes $\hat{f}_0$ for the ensuing processing cycle, thereby guaranteeing temporal continuity and robustness in long term force estimation. All experiments were trained for 50 epochs with a batch size of 128, utilizing the AdamW optimizer and a learning rate of $1 \times 10^{-4}$. During inference on an NVIDIA RTX 4090, our FGER-Net model achieves an latency of 0.87 ms, with a computational cost of 2.0 GFLOPs. The model comprises a total of 0.517 million parameters.

\section{EXPERIMENT}
\IEEEpubidadjcol

\subsection{Sensing Vibrations}

\subsubsection{Experimental Setup}

A high frequency vibration experiment was designed to evaluate the vibration sensing capability of Mixtac. It was used to test whether the sensor can capture dynamic vibration signals for effective fast adapting input to the fusion network.\par

\begin{figure}[htbp]
    \centering
    \includegraphics[width=3.5in]{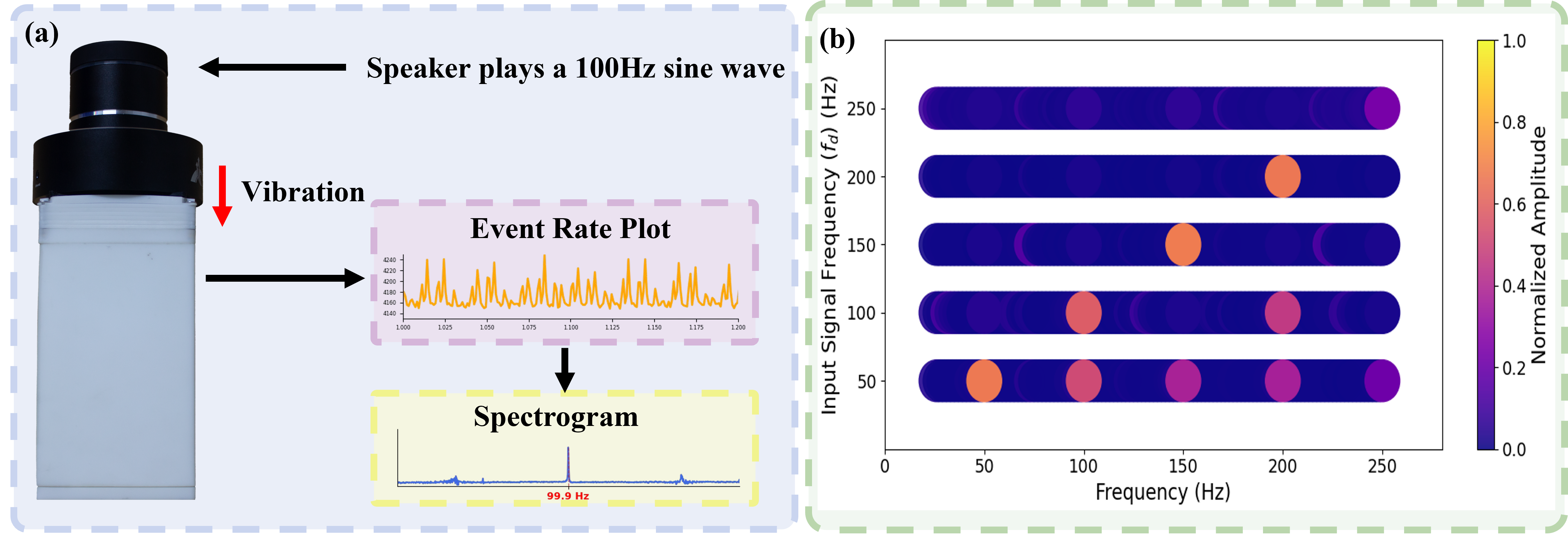}
    \caption{Experimental setup and results for high-frequency vibration sensing. (a) A resonant Bluetooth speaker (Adin S9WiFi) is placed directly on the Mixtac sensor to act as a vibration source. (b) Normalized frequency spectrum obtained from Mixtac's raw event stream under various input tones. The spectrum was generated by applying a Fourier transform to the event count time series. The resulting prominent diagonal pattern indicates a strong linear correlation, confirming that the reconstructed frequency consistently matches the speaker's input frequency up to the tested maximum of 250 Hz.}
    \label{zhendongshiyan}
\end{figure}

Following Funk et al.~\cite{funk2024evetac} , a resonant speaker (Adin S9WiFi) is placed in direct contact with the silicone surface to maximize vibration energy transfer (Fig. \ref{zhendongshiyan}(a)). Sinusoidal audio stimuli ranging from 50 to 250 Hz are applied. To process the data, raw asynchronous events are binned into 2 ms windows, generating a 500 Hz synchronous event rate signal.\par

\subsubsection{Result}

The results confirm that Mixtac can reconstruct high frequency vibrations from its raw event stream. The reconstructed dominant frequency remains in close agreement with the speaker input frequency over 50 to 250 Hz in (Fig. \ref{zhendongshiyan}(b)). This range is close to the Nyquist limit of the 500 Hz acquisition rate.\par

\subsection{Stability of Long Term Force Estimation}
\subsubsection{Experimental Setup}
This experiment examines whether the model remains stable during sustained contact. To reveal long term drift, the sensor was pressed for about 25 s with an 18 mm hemispherical 3D printed indenter while mounted on a 6 axis force sensor operating at 800 Hz. Stability was quantified by the MAE between the predicted and ground truth normal force ($F_z$). A 0.5 s sliding window was used to obtain rolling MAE and track error growth over time. Results from 10 trials were summarized in 5 s bins with violin plots. The dataset was split at the trajectory level with an 8:2 training validation ratio to prevent data leakage.
\begin{figure}[htbp] 
  \centering
  \includegraphics[width=3.5in]{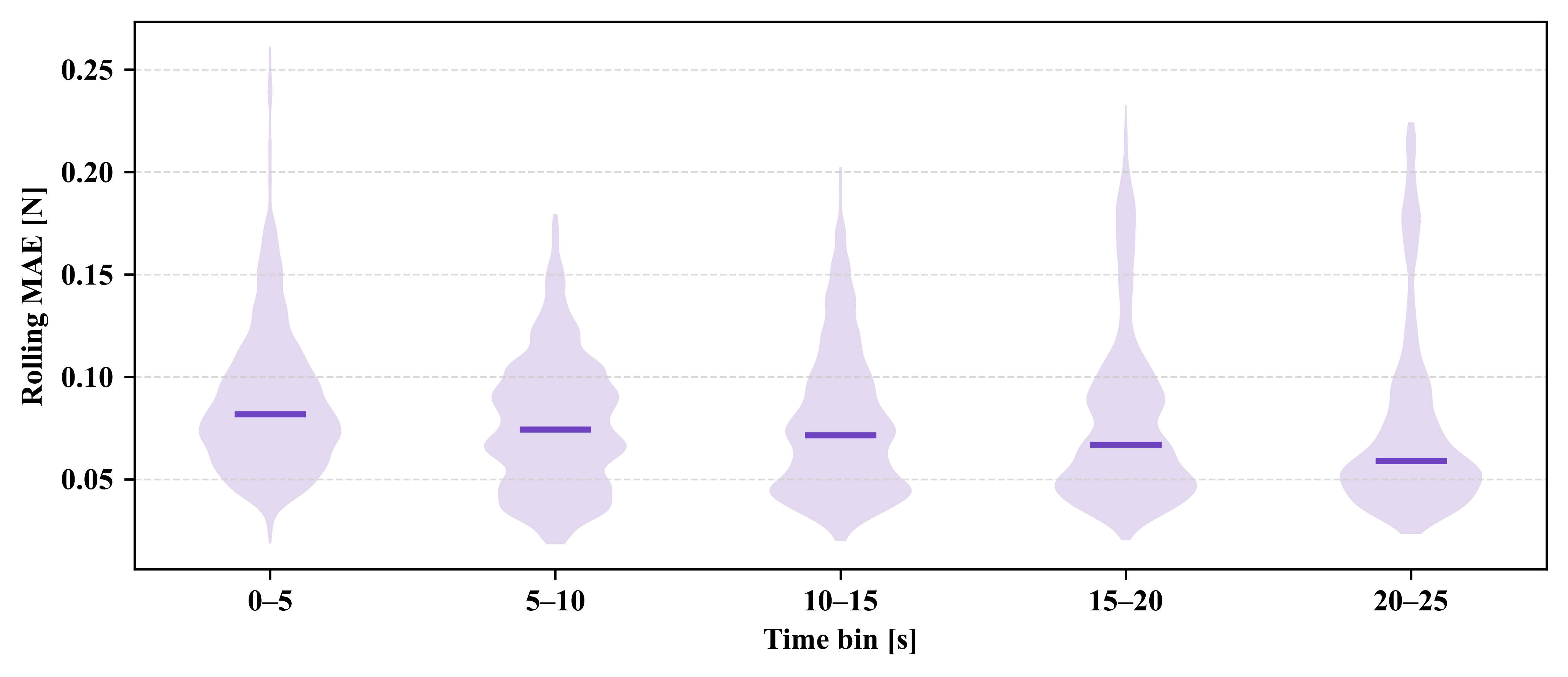}
  \caption{Time-binned violin plot of rolling MAE for normal-force estimation demonstrates that the error distribution remains stable (and slightly decreases) over the trial duration. Rolling MAE is computed on $F_z$ using a 0.5 s sliding window. Each violin aggregates rolling MAE samples from ten trials, pooled within 5 s bins (0 to 5, 5 to 10, 10 to 15, 15 to 20, 20 to 25 s); the horizontal bar denotes the median within each bin.}
  \label{fig_long}
\end{figure}

\subsubsection{Result}

The long term stability of normal force estimation is characterized by time-binned violin plots of the rolling MAE aggregated over multiple trials (Fig. \ref{fig_long}). The distributions remain stable across time, with median rolling MAE values for the five bins of 0.08N (0-5s), 0.07N (5-10s), 0.07N (10-15s), 0.07N (15-20s), and 0.06N (20-25s).

\subsection{Performance Comparison of Different Input Signals}

\subsubsection{Experimental Setup}

To demonstrate the essential synergy between frame and event data, a comparative experiment was conducted. The primary objective was to quantify the performance improvement gained from this synergistic fusion by comparing the hybrid system against baselines that rely on only a single modality.

For this comparison, three distinct model architectures were evaluated, each designed to isolate the different input signals:
\begin{enumerate}

    \item \textbf{Frames + Events Model}: The complete FGER-Net as proposed in this paper.

    \item \textbf{Frames only Model}: An ablated model was built without the event recurrent branch. The frame feature was fed directly into a regression head to isolate the contribution of spatial information.

    \item \textbf{Events only Model}: An ablated model was built without the frame guided correction module. Its force estimation relies only on recurrent event integration to isolate the contribution of temporal dynamics without absolute correction.
\end{enumerate}

AAll three models were trained under the same protocol and evaluated on dynamic force profiles. Performance was assessed by comparing predicted and ground truth force trajectories, with MAE and RMSE used as quantitative metrics for normal force estimation.

\subsubsection{Result}
\begin{figure*}[htbp]
    \centering
    \includegraphics[width=7.16in]{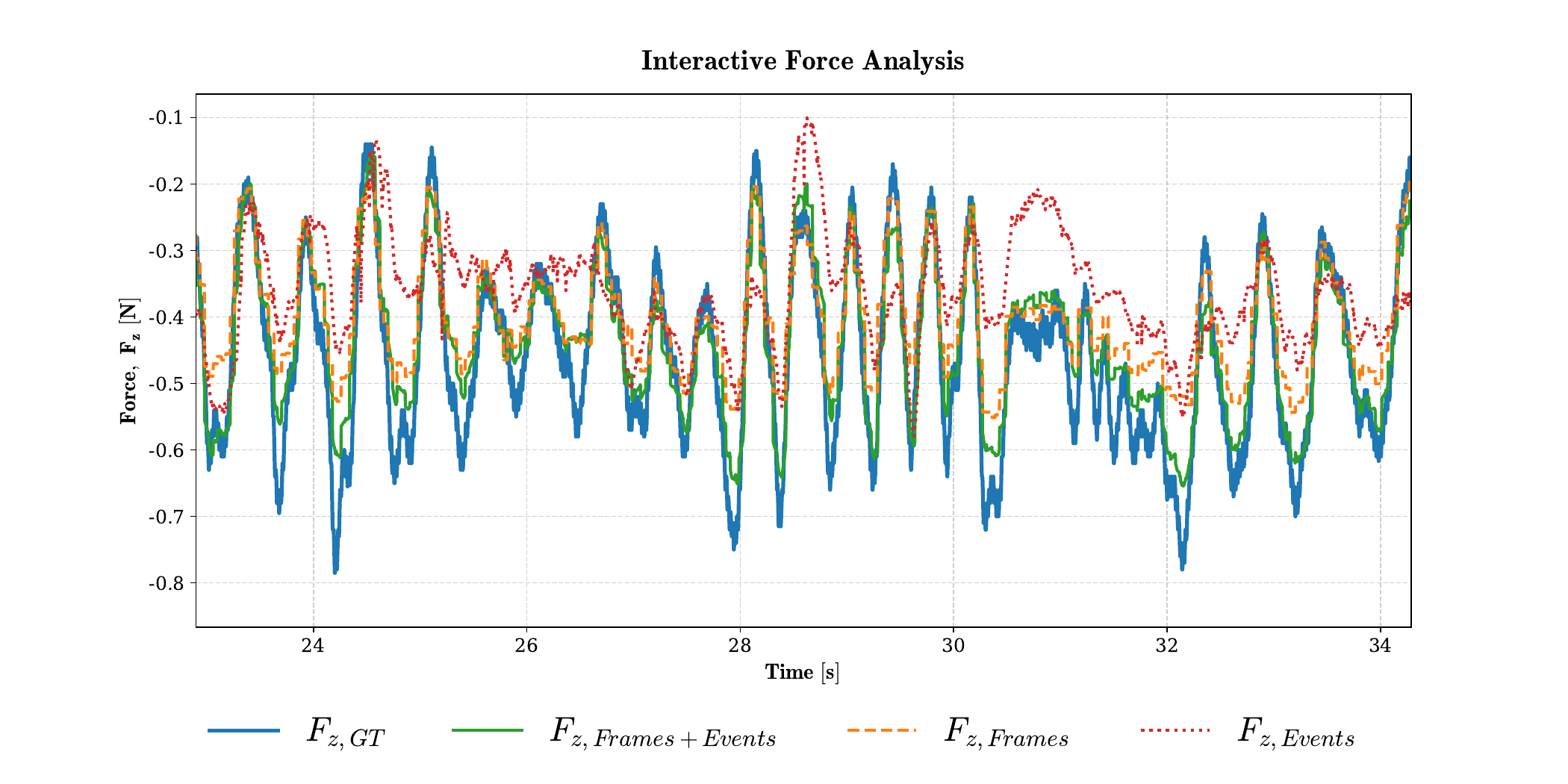}
    \caption{Results of different input signals. The predicted normal force is compared against the ground truth (blue) for three input modalities. The proposed frames+events model (green) accurately tracks high-frequency dynamics. In contrast, the frames only model (orange) misses fine details due to its low update rate, while the events only model (red) suffers from significant drift. The complementary nature of frames and events is thus demonstrated, validating that their fusion is essential for achieving both high frequency response and long term stability.}
    \label{diff_signals}
\end{figure*}

Three models were evaluated to compare input modalities: Frames only, Events only, and the proposed FGER-Net (Fig.~\ref{diff_signals}, Table~\ref{tab:inputs}). The Frames-only model provided a stable baseline but underrepresented high-frequency dynamics (MAE/RMSE: 0.06/0.08 N), whereas the Events-only model captured rapid oscillations, exhibited baseline drift, resulting in the largest error (0.12/0.14 N). The fused model achieved both high frequency tracking and baseline stability, yielding the best accuracy (0.04/0.05 N).\par

\begin{table}[htbp]
\caption{Comparison of Different Input Signals for Force Estimation\label{tab:inputs}}
\centering
\begin{tabular}{lcccc}
\hline
          &  &  & \textbf{Long Time} & \textbf{Sampling} \\
\textbf{Input Signals} & \textbf{MAE} & \textbf{RMSE} & \textbf{Force} & \textbf{Rate} \\
          & \textbf{(N)} & \textbf{(N)} & \textbf{Estimation} & \textbf{(Hz)} \\
\hline
Frames        & 0.06 & 0.08 & $\checkmark$ & 25  \\
Events        & 0.12 & 0.14 & $\times$     & 500 \\
Frames+Events & \textbf{0.04} & \textbf{0.05} & $\checkmark$ & 500 \\
\hline 
\end{tabular}
\end{table}
\subsection{Performance Comparison of Different Fusion Architectures}

\subsubsection{Experimental Setup}

To validate the effectiveness of our proposed architecture, a comparative experiment was conducted. This experiment was designed to demonstrate that the specific method of combining event and frame information is critical for achieving high performance, and that feature fusion strategy is insufficient for capturing high frequency dynamics.\par

To this end, our proposed FGER-Net was directly compared against a common baseline fusion strategy, referred as the feature level deep fusion model.\par

\begin{enumerate}

\item \textbf{Feature Level Deep Fusion Model (Baseline)}: Operating in parallel, this model aggregates inter frame events into sparse frames, applying sparse convolutions exclusively to non zero pixels. Two independent CNN backbones extract features from both modalities, which are subsequently channel concatenated and regressed into force estimates via an MLP.

\item \textbf{FGER-Net (Proposed)}: Diverging from naive parallel concatenation, our architecture employs a sequential corrective mechanism: it continuously integrates high frequency force increments from the event stream while periodically correcting these accumulated estimates using absolute image frames.
\end{enumerate}

Both architectures were trained and evaluated on the same static force dataset. Performance was measured by MAE, RMSE, and $R^2$ against the ground truth. The dataset contains 30 trials of 25 s each, generated by applying forces from 0 to 1 N with a 3D printed contactor. It was split into training and validation sets at an 8:2 ratio at the trajectory level to prevent data leakage from temporally correlated samples.\par

\subsubsection{Result}
\begin{figure}[htbp]
\centering
\includegraphics[width=3.5in]{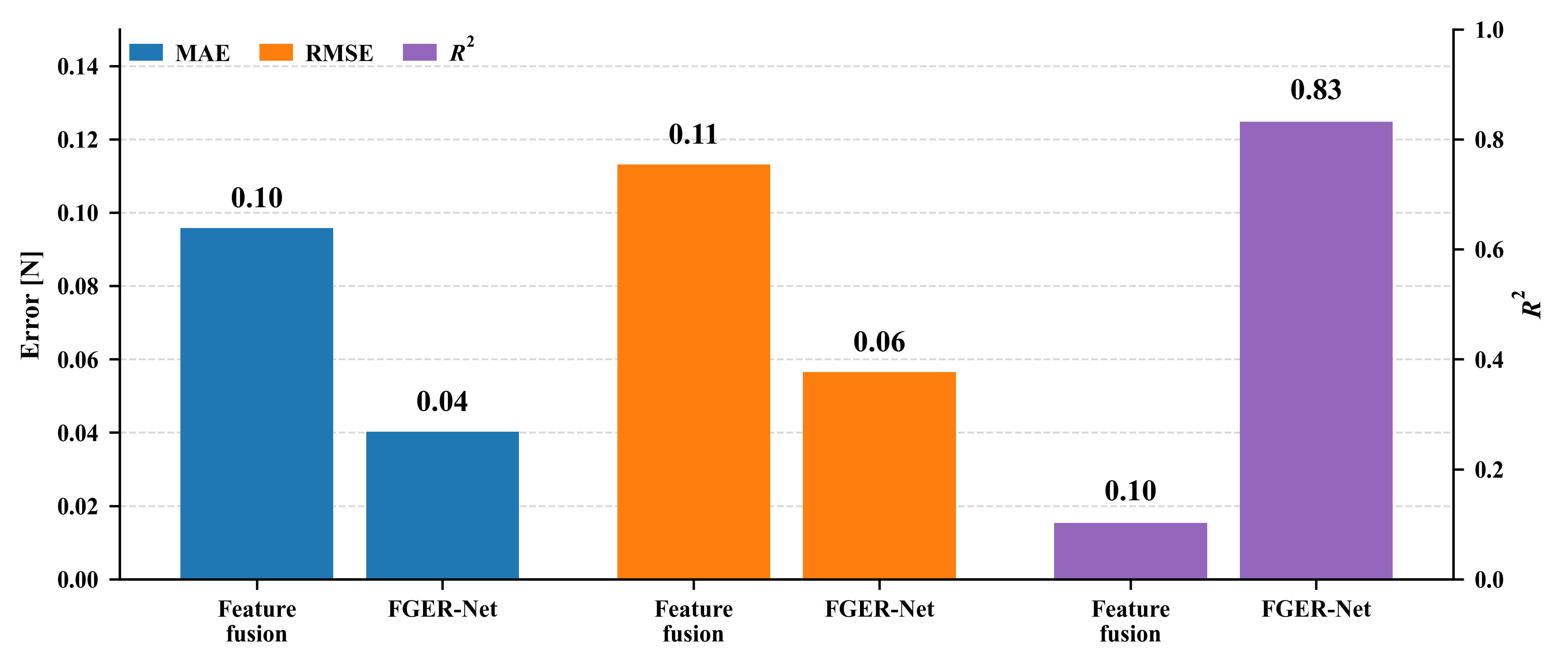}
\caption{Performance comparison for normal force estimation between the proposed FGER-Net and a baseline feature level fusion model, evaluated on MAE, RMSE, and $R^2$ metrics. The FGER-Net yields lower error metrics (MAE and RMSE) and a higher $R^2$ than the feature level fusion model, indicating a closer fit to the ground truth data. }
\label{diff_model}
\end{figure}

The experimental results show a difference in performance between the two models in estimating the normal force (Fig. \ref{diff_model}). The proposed FGER-Net had lower error metrics than the baseline feature fusion model. FGER-Net's MAE was 0.04 N and its RMSE was 0.05 N, compared to the feature fusion model's MAE of 0.10 N and RMSE of 0.11 N. In terms of the $R^2$ value, FGER-Net achieved 0.83, while the baseline model achieved 0.10.
\subsection{Tracking Hybrid Force Profile of Transient Impact and Static Hold}

\subsubsection{Experimental Setup}
To validate FGER-Net during dynamic-to-static transitions, a hybrid force profile was evaluated using a controlled impact test. A calibrated 20 g cylindrical metal weight was dropped from a fixed height of 5 cm onto the sensor center. Across 30 trials, the full contact sequence, including the initial impact, subsequent bounces, and final static hold, was synchronously recorded with 6-axis ground-truth force readings. To avoid data leakage, the dataset was split into training and validation sets at an 8:2 ratio at the trajectory level.\par

\subsubsection{Result}
\begin{figure}[htbp]
    \centering
    \includegraphics[width=3.5in]{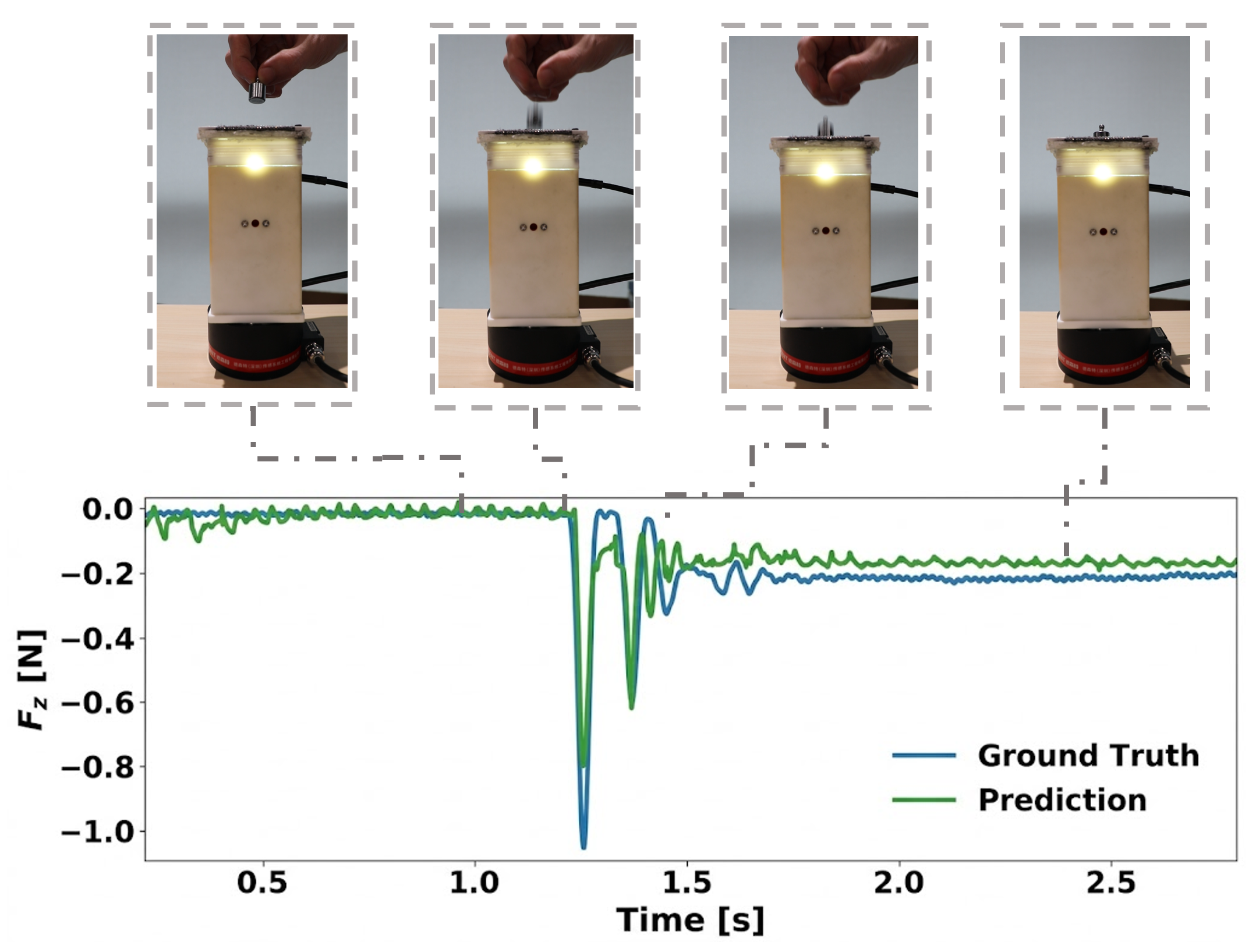}
    \caption{Results for tracking a hybrid force profile. A 20 g cylindrical weight was dropped onto the sensor to generate a force signal with both transient impact and sustained static force. The plot compares the normal force predicted by FGER-Net (green) with the ground truth measurement (blue). The prediction remains consistent with the ground truth during both the impact and static hold phases (MAE = 0.05 N, RMSE = 0.07 N). This result supports the proposed fusion architecture, which combines event driven dynamic tracking with frame guided static estimation.}
    \label{dongtai}
\end{figure}

The results of this case study unequivocally demonstrate Mixtac's unique ability to robustly handle complex force profiles by leveraging its hybrid sensing architecture. The normal force profile predicted by our model seamlessly tracks the ground truth throughout both the transient impact and the ensuing static hold phases, indicating a flawless transition between the event driven and frame guided sensing regimes. 
details of the impact dynamics. These details include not only the primary impact peak but also the secondary peak resulting from a subsequent mechanical bounce. The predicted force curve, generated by the 20 g falling weight, accurately reproduces the actual physical force profile, achieving an MAE of 0.05 N and an RMSE of 0.07 N (Fig. \ref{dongtai}).

\subsection{Closed loop Slip Control Experiment}

\begin{figure}[htbp]
\centering

\includegraphics[width=3.5in]{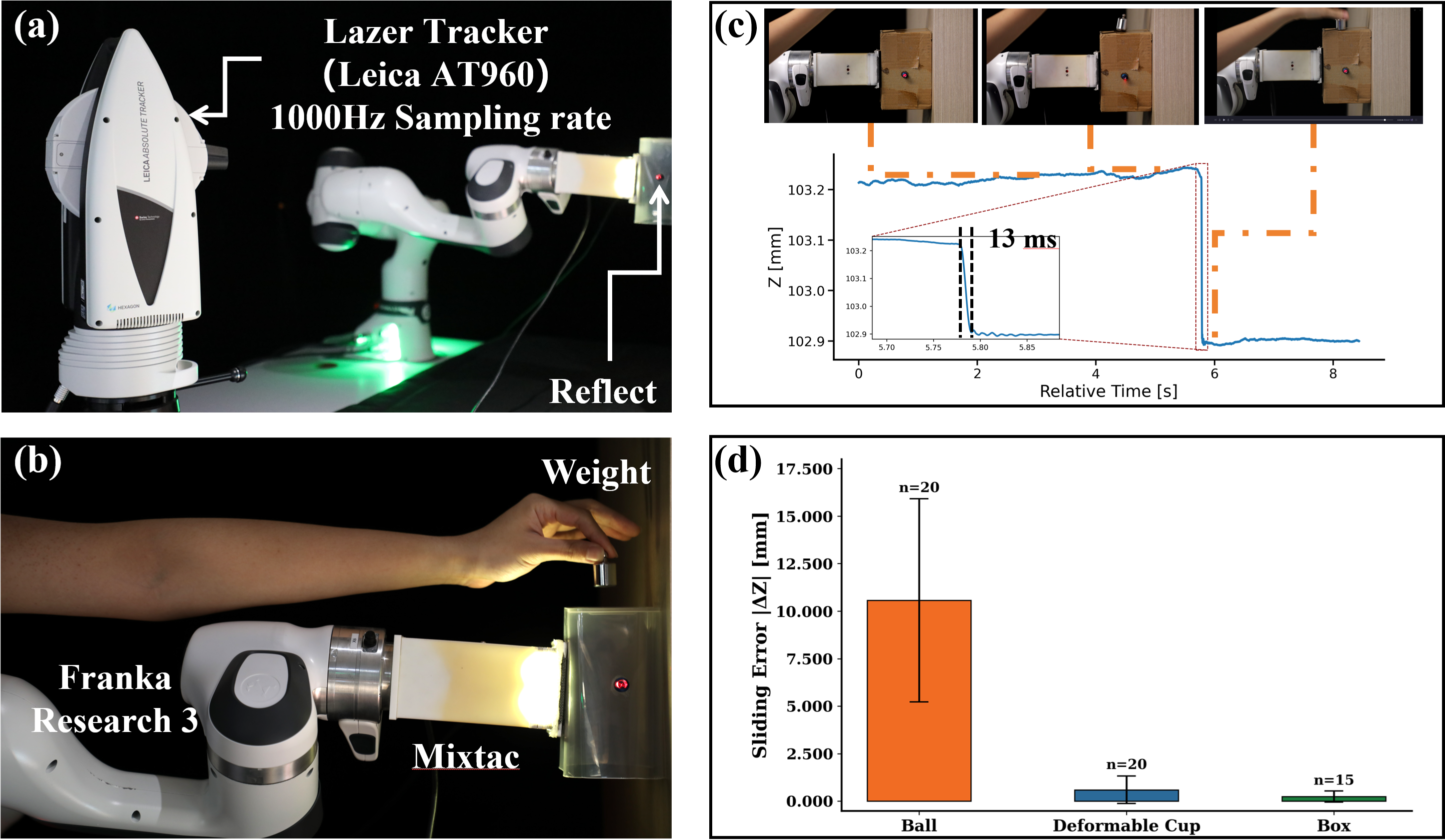}

\caption{Closed loop slip control experiment. (a) The test object, marked with a retroreflective target, is pressed against a wall. A laser tracker (Leica AT960; $1000~\text{Hz}$, $\pm3~\mu\text{m}$) provides the ground truth slip displacement. A $100~\text{g}$ weight is dropped from $5~\text{cm}$ to introduce a sudden disturbance. (b) The weight is released above the object and falls freely. (c) Three keyframes from the experiment and the corresponding slip displacement curve. From slip onset to control completion, the process lasts $13~\text{ms}$ and produces a slip of $0.32~\text{mm}$. (d) Sliding error across different test objects.}
\label{slip}
\end{figure}

\subsubsection{Experimental Setup}

To evaluate the real time capability of Mixtac hardware, a closed loop slip control task was performed with a Franka Research 3 robotic arm. The end effector pressed diverse objects against a vertical wall at the critical force required to prevent slip. External disturbances from calibrated weights consistently induced slip without active intervention, providing a baseline for evaluating algorithm sensitivity and responsiveness.\par

The closed loop controller uses a proportional response driven by real time event density. Slip is detected when the 1\,ms event count ($N_t$) exceeds the background threshold ($N_{\mathrm{th}}$). The arm then advances by $\Delta d_t = k \cdot (N_t - N_{\mathrm{th}})$, increasing the normal force on the object.\par

Furthermore, to evaluate the system's robustness across distinct physical properties, three everyday objects (a ball, a deformable cup, and a box) were selected, representing a diverse range of materials, shapes, and weights. All displacement measurements were obtained using a high precision laser tracker, which provided ground truth data for object movement~(Fig. \ref{slip}(a)).\par

For quantitative evaluation, the relative displacement of objects from their initial positions was measured to assess the precision of our perception system.

\subsubsection{Result}
During the critical slip state, the box was impacted by a 100 g mass freely dropped from a height of 5 cm (Fig.~\ref{slip}(a)). Subsequently, sliding commenced, and within 13 ms of slip onset, the slip control algorithm was activated and successfully prevented further downward motion of the box (Fig.~\ref{slip}(b)).

Beyond this representative box trial, control performance was further evaluated across objects with different physical properties. The box-trial keyframes and corresponding slip displacement curve are presented (Fig.~\ref{slip}(c)). The sliding errors across different test objects are summarized (Fig.~\ref{slip}(d)). Twenty trials were conducted for the rigid ball and deformable cup, and 15 for the box. The mean sliding displacement errors were 0.90 mm for the cup, 0.71 mm for the box, and 10.03 mm for the ball. The larger error for the ball is expected because its rigid spherical geometry reduces the contact area and available friction. Even so, the controller stabilized all objects in all trials, highlighting the ultra-low-latency event-driven reflex.

\section{DISCUSSION}

\subsection{Analyzing Experimental Results}

This work addresses a central tradeoff in vision based tactile sensing. It also presents a dedicated hybrid tactile hardware prototype for this purpose. A single sensor should support both rapid dynamic perception and stable long term force estimation within the same contact episode. The discussion below revisits this tradeoff through the experiments.\par

The high frequency vibration reconstruction experiment first clarifies why a hybrid design is necessary. Mixtac reconstructs vibrations up to 250 Hz from raw event counts. This result shows that the hardware can preserve rapid temporal changes. Such changes are important for dynamic tactile perception. However, temporal responsiveness alone is not enough. A tactile sensor for manipulation also remain reliable after contact becomes steady. This is the regime in which purely event driven sensing often begins to drift.\par

The long duration static force estimation experiment addresses this second requirement. Mixtac maintains static force estimation over 25 s. Robotic contact rarely ends at impact. Many tasks proceed from approach, to contact, and then to hold. A sensor that responds only to change cannot support this full sequence. In the proposed system, frame signals provide an absolute spatial reference. This reference reduces the drift that often limits event only integration during prolonged contact.\par

The ablation and baseline comparison experiments further clarify the source of improvement. The gain does not come from simply adding another input stream. Frames only estimation remains stable, but it misses rapid transients. Events only estimation responds quickly, but its estimates drift over time. Naive feature fusion is also insufficient. Dense frame features can dominate training. As a result, they can suppress the temporal contribution of events. For this reason, the structured design of FGER-Net is as important as the hybrid hardware itself. The architecture allows each modality to contribute where it is most informative. It does not force an unbalanced compromise.\par

The hybrid force profile experiment provides the clearest evidence for this transition. The sequence begins with a transient impact and ends with a static hold. In the early stage, the event stream captures the rapid force variation, including the impact peak and the rebound. Once the contact settles, the frame stream becomes more important and maintains a stable estimate. The two modalities therefore support different phases of the same tactile process. This result supports the bio inspired motivation introduced earlier.\par

The closed loop slip control experiment further reveals the capability of the hardware for low latency tactile sensing. The focus here is early signal availability under rapid disturbance. The 13 ms reaction latency shows that the event pathway can detect slip related changes before large failure develops. This result supports the potential of the hardware for low latency tactile feedback. It also supports the broader argument of the paper. Dynamic responsiveness and steady estimation belong to the same sensing requirement in robotic interaction.\par

Taken together, these experiments address the challenge posed in the Introduction. Mixtac relaxes the classical tradeoff through cooperation between frame and event sensing. The system achieves both high frequency tracking and stable long term estimation. This result establishes a practical path toward vision based tactile sensing that is both responsive and stable. It also indicates potential for robotic manipulation tasks that require fast response and sustained contact estimation.\par

\subsection{Signal Considerations}

\begin{table}[!t]
\caption{Comparison of Typical Vision Based Tactile Sensors\label{tab:comparison}}
\centering

\resizebox{\columnwidth}{!}{
\begin{tabular}{l c c c c}
\hline 
 & & \textbf{Sampling} & \textbf{Normal} & \textbf{Normal} \\
 \textbf{Sensor}& \textbf{Input} & \textbf{Rate} & \textbf{Force Error} & \textbf{Force Range} \\
 & \textbf{Signal} & \textbf{(Hz)} & \textbf{(N)} & \textbf{(N)}    \\
\hline 
Insight\cite{sun2022soft}          & Frames        & 40    & 0.03 & 2 \\ 
ViTacTip\cite{zhang2025design}     & Frames        & 30    & 0.04 & $\sim$1 \\
E-BTS\cite{mukashevEBTSEventBasedTactile2025}          & Events        & 500  & 0.49 & 8 \\
Gelevent\cite{yin2025gelevent}      & Events        & 180   & 0.80 & 16 \\

\textbf{\textit{Mixtac} (ours)} & Frames+Events & \textbf{500}   & 
\textbf{0.04} & \textbf{1} \\

\hline 
\end{tabular}
}
\end{table}

The performance of the Mixtac sensor was benchmarked against other vision based tactile sensors, with the results summarized in Table~\ref{tab:comparison}. Our findings show that Mixtac achieves a normal force range and error level comparable to mainstream vision based tactile sensors. Furthermore, compared to purely event-based approaches like~\cite{mukashevEBTSEventBasedTactile2025}, which rely on active high frequency illumination to artificially induce events for static perception, Mixtac leverages the synergistic nature of frames. This allows for stable long term estimation while preserving the intrinsic sparsity and bandwidth efficiency of event data, avoiding the high data redundancy caused by flickering light sources.\par

\IEEEpubidadjcol
\subsection{Limitations and Future Work}

Despite performance, Mixtac remains constrained by deployability and synchronization. Its current form factor, dominated by the DAVIS346, limits integration into grippers and dexterous hands, although fingertip-scale event cameras (e.g., PX EVB Gen2 and Speck) provide a clear miniaturization path. In addition, residual latency between force ground truth and event streams still bottlenecks estimation accuracy, motivating a custom PCB for hardware time synchronization.

\par

Future work will focus on the following key areas. The first will be to address the hysteresis effect and low intrinsic frequency of silicone in order to restore high frequency vibration signals for real time texture recognition. Moreover, one promising avenue for future research lies in integrating our tactile perception system with visual modalities to facilitate multimodal dataset construction.

\par

\IEEEpubidadjcol
\IEEEpubidadjcol

\section{CONCLUSION}
\IEEEpubidadjcol

This paper introduced Mixtac, a novel hybrid tactile sensor designed to resolve the fundamental tradeoff between high frequency dynamic sensing and long term estimation stability. FGER-Net was developed to synergistically fuse the two data streams. In this architecture, frames are used to periodically correct for drift during training and to continuously stabilize the high frequency force predictions derived from the event stream during inference. Our experiments validate this approach, demonstrating a low MAE of 0.04 N and significant performance gains over single modality methods. By achieving both high fidelity dynamic force tracking and stable force estimation over extended periods, this work paves the way for more complex and robust robotic manipulation in unstructured environments. \par


\bibliographystyle{Bibliography/IEEEtranTIE}
\bibliography{sum}\ 

\begin{IEEEbiography}[{\includegraphics[width=1in,height=1.25in,clip,keepaspectratio]{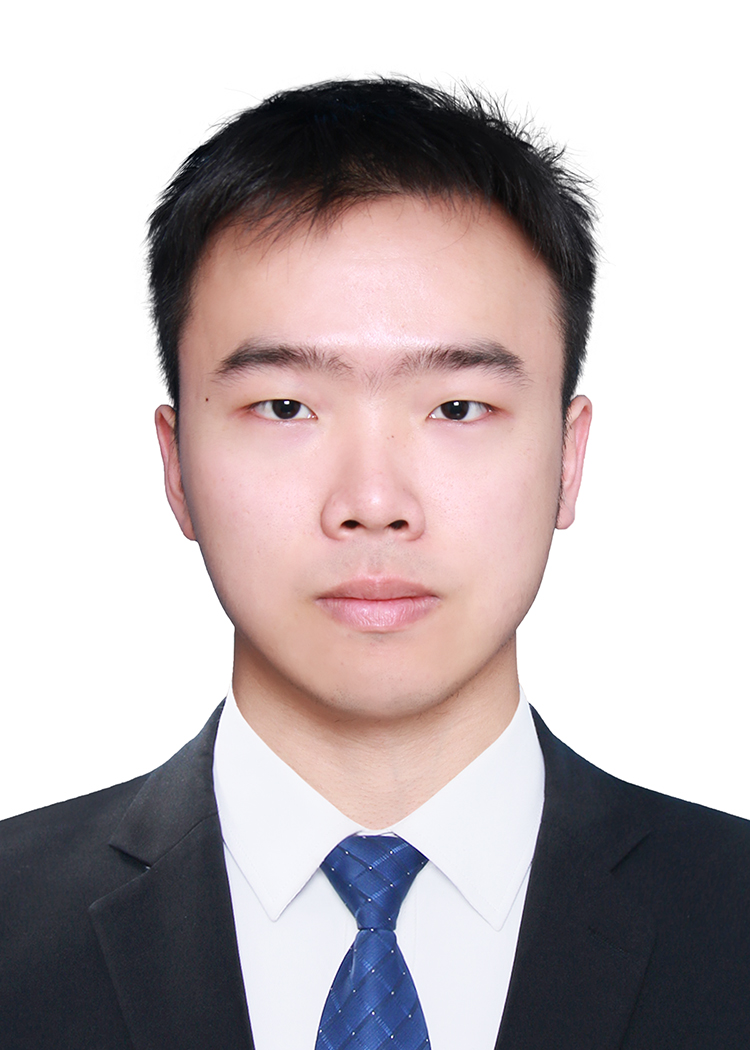}}]{Yihang Li}
received the B.E. degree in robotics engineering from Harbin Engineering University, in 2025, China. He is currently working toward the Ph.D. degree at the Shanghai Research Institute for Intelligent Autonomous Systems, Tongji University, China. 

His research interests include tactile sensors, intelligent sensing, and robotics.
\end{IEEEbiography}

\begin{IEEEbiography}[{\includegraphics[width=1in,height=1.25in,clip,keepaspectratio]{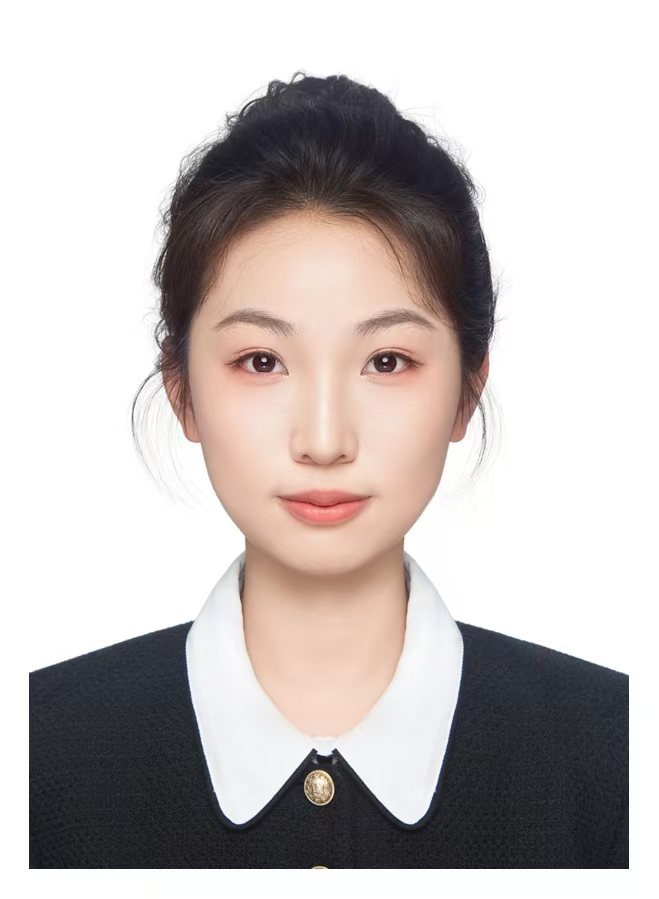}}]{Yijin Chen} is with the Robotics Engineering program at the Guohao Academy of Future Technology, Tongji University. She serves as a Research Assistant at the State Key Laboratory of Autonomous Intelligent Unmanned Systems. Her research interests include robotics system design, intelligent control, and human-robot interaction.
\end{IEEEbiography}

\begin{IEEEbiography}[{\includegraphics[width=1in,height=1.25in,clip,keepaspectratio]{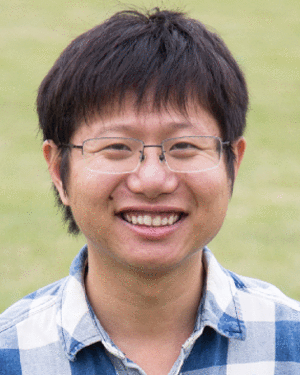}}]{Junkai Xu} received the B.S. degree from Nanchang University, China, in 2011, and the M.S. degree from Shandong University, China, in 2014. From 2014 to 2020, he studied as a Ph.D. candidate at Shanghai Jiao Tong University, China. He is with IMU Master Technology Co., Ltd, Shanghai 200240, China, focusing on embedded system development, hardware-software co-debugging, and performance optimization.
\end{IEEEbiography}

\begin{IEEEbiography}[{\includegraphics[width=1in,height=1.25in,clip,keepaspectratio]{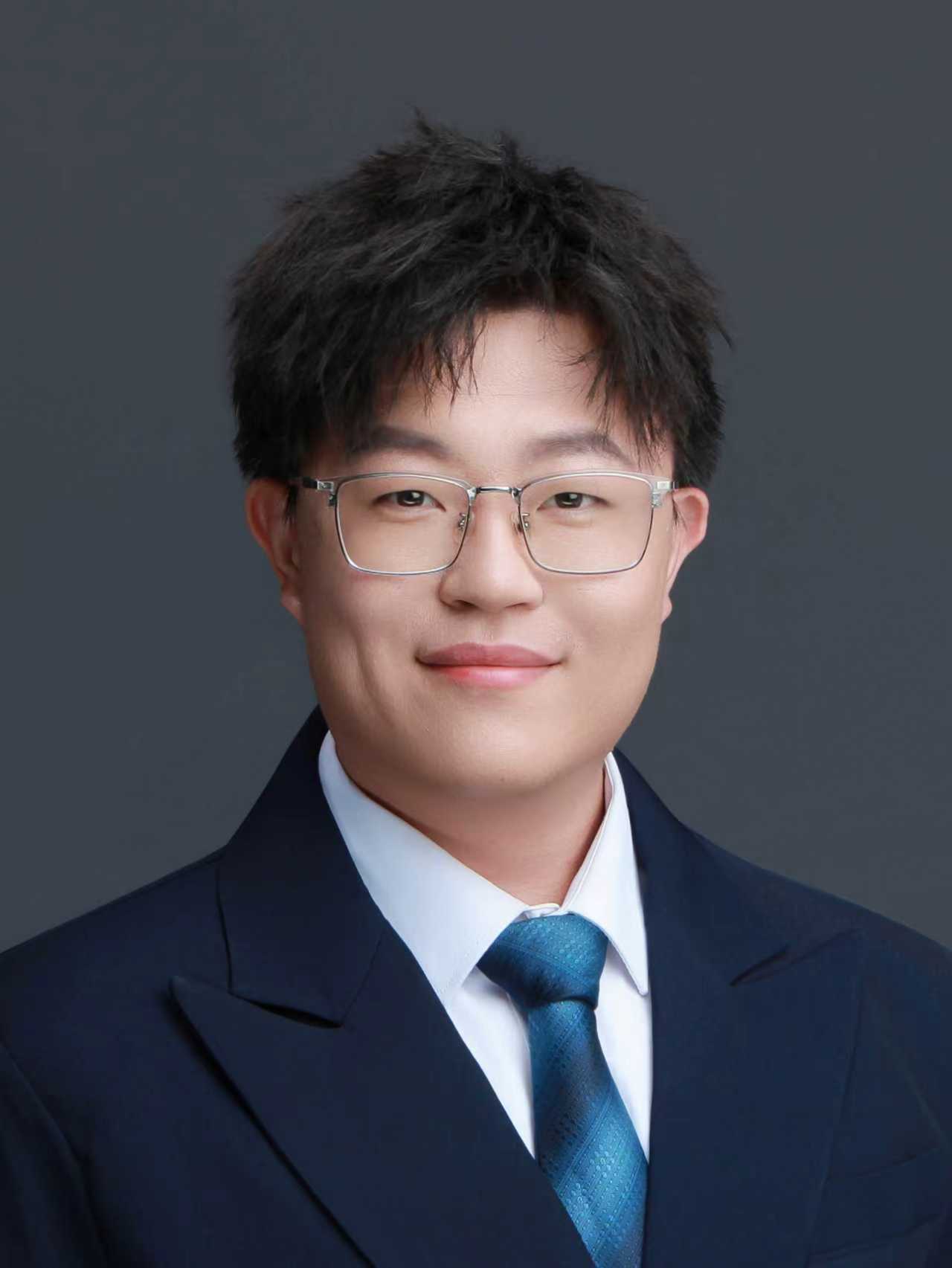}}]{Na Ningguta}
received the B.E. degree from Northeast Forestry University, China. He is currently an Embedded Software Engineer with Shanghai Guanshi Technology Co., Ltd, focusing on embedded system development, hardware-software co-debugging, and performance optimization.
\end{IEEEbiography}

\begin{IEEEbiography}[{\includegraphics[width=1in,height=1.25in,clip,keepaspectratio]{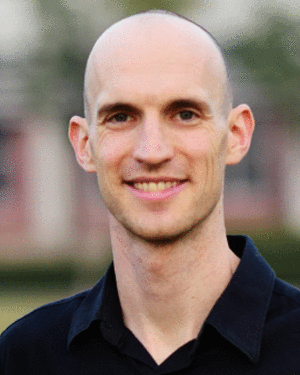}}]{Peter B. Shull}
(Member, IEEE) received the B.S. degree in mechanical engineering and computer engineering from LeTourneau University, Longview, TX, USA, in 2005, and the M.S. and Ph.D. degrees in mechanical engineering from Stanford University, Stanford, CA, USA, in 2008 and 2012, respectively.
From 2012 to 2013, he was a Postdoctoral Fellow with the Bioengineering Department, Stanford University. 

He is currently a Professor of Mechanical Engineering with Shanghai Jiao Tong University, Shanghai, China. He has conducted pioneering research in human-computer interaction, hand gesture recognition, wearable systems.
\end{IEEEbiography}

\begin{IEEEbiography}[{\includegraphics[width=1in,height=1.25in,clip,keepaspectratio]{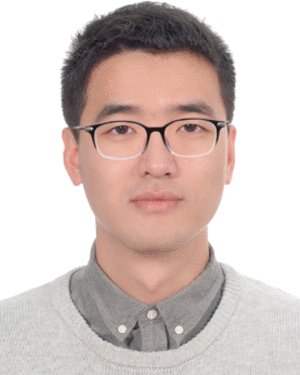}}]{Shuo Jiang}
(Member, IEEE) received the B.E. degree in mechatronic engineering from Zhejiang University, Hangzhou, China, in 2015, and the Ph.D. degree in mechanical engineering from Shanghai Jiao Tong University, Shanghai, China, in 2020. From 2019 to 2020, he was a Visiting Scholar with Imperial College London, London, U.K. 

He is currently an Associate Professor with the Department of Control Science and Engineering, College of Electronics and Information Engineering, Tongji University, Shanghai. His research interests include human machine interaction, intelligent sensing, and robotics.
\end{IEEEbiography}
 
\begin{IEEEbiography}[{\includegraphics[width=1in,height=1.25in,clip,keepaspectratio]{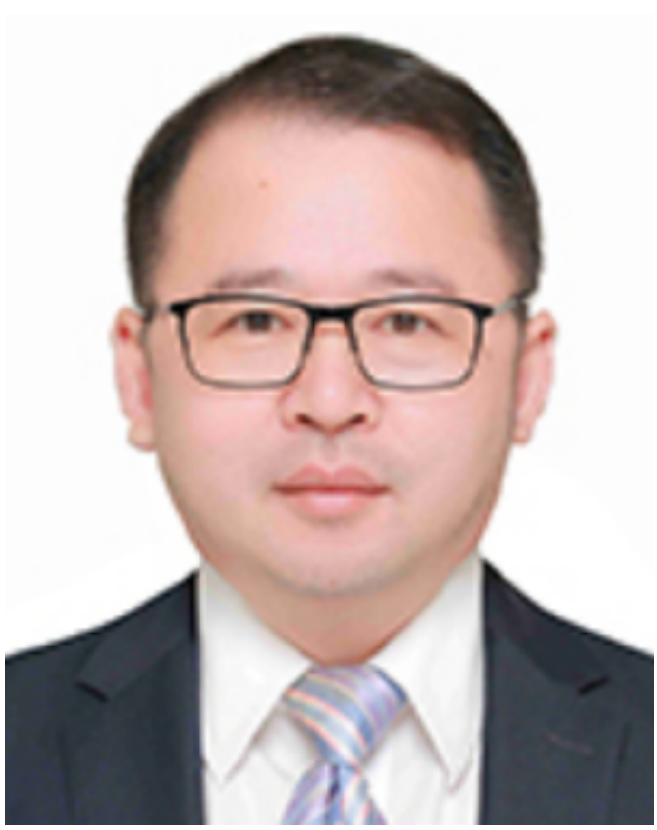}}]{Bin He}
(Senior Member, IEEE) received the B.S. degree in engineering machinery from Jilin University, Changchun, China, in 1996, and the Ph.D. degree in mechanical and electronic control engineering from Zhejiang University, Hangzhou, China, in 2001. From 2001 and 2003, he was a Postdoctoral Researcher with the State Key Laboratory of Fluid Power Transmission and Control, Zhejiang University.

He is currently a Professor with the Department of Control Science and Engineering, College of Electronics and Information Engineering, Tongji University, Shanghai, China. His current research interests include intelligent robot control, biomimetic microrobots, and wireless networks.
\end{IEEEbiography}

\end{document}